\documentclass[letterpaper]{article} 
\usepackage{aaai2026}  
\usepackage{times}  
\usepackage{helvet}  
\usepackage{courier}  
\usepackage[hyphens]{url}  
\usepackage{graphicx} 
\usepackage{natbib}  
\usepackage{caption} 
\usepackage{algorithm}
\usepackage{algorithmic}
\usepackage{array}

\usepackage{placeins}
\usepackage{booktabs}
\usepackage{subcaption}
\usepackage{tikz}
\usetikzlibrary{arrows.meta,positioning}
\usepackage{amsmath}
\usepackage{xcolor}

\usepackage{tabularx}

\usepackage{booktabs}
\usepackage{graphicx}

\usepackage[most]{tcolorbox}
\definecolor{promptbg}{HTML}{F8F9FA}
\definecolor{promptframe}{HTML}{E2E8F0}
\definecolor{varcolor}{HTML}{0D9488}
\definecolor{rolecolor}{HTML}{1E293B}

\newcommand{\pvar}[1]{\textcolor{varcolor}{\texttt{\{\{#1\}\}}}}

\newtcolorbox[auto counter]{promptbox}[2][]{%
  enhanced,
  breakable,
  colback=promptbg,
  colframe=promptframe,
  arc=2mm,
  boxrule=0.8pt,
  fontupper=\small\sffamily,
  title=\textbf{Prompt A.\thetcbcounter: #2},
  coltitle=rolecolor,
  colbacktitle=promptframe!40!white,
  attach boxed title to top left={yshift=-2mm, xshift=2mm},
  boxed title style={boxrule=0pt, colframe=white, arc=1mm},
  #1
}

\usepackage{newfloat}
\usepackage{listings}
\DeclareCaptionStyle{ruled}{labelfont=normalfont,labelsep=colon,strut=off} 
\floatstyle{ruled}
\newfloat{listing}{tb}{lst}{}
\floatname{listing}{Listing}
\title{Who Belongs Together? Topical and Social Structure in Bluesky Starter Packs}

\author{
    Sima Adleyba\textsuperscript{\rm 1},
    Onur Varol\textsuperscript{\rm 1,2}
}

\affiliations{
\textsuperscript{\rm 1} Faculty of Engineering and Natural Sciences, Sabanci University, Istanbul, Türkiye \\
\textsuperscript{\rm 2} Center of Excellence in Data Analytics, Sabanci University, Istanbul, Türkiye
}

\usepackage{bibentry}

\usepackage{xcolor}

\begin{document}

\maketitle

\begin{abstract}
Bluesky starter packs are human curated collections of accounts and feeds aimed at helping users discover new communities, especially during onboarding. Prior research has examined their effects on platform growth and account visibility. However, whether these packs capture meaningful topical and social structure remains unanswered. We study 2,718 active starter packs containing $\approx144K$ accounts, combining pack metadata, follow graphs, user post histories, and inferred demographic attributes to answer this question. We assign starter packs to 17 topical categories, and examine semantic coherence, curator--member resemblance, shared audiences, and follow reciprocity. 
We found that pack members are closely aligned with both their fellow pack members and their pack's description compared to accounts included in other packs of the same topic, indicating topical specificity beyond the broader topical categories. We also identify semantic resemblance between curators and their pack members, while demographic resemblance varies by attribute and topic. Network analyses show higher follow reciprocity among members of the same pack and between members and curators than between a pack member and either another member from the same broader topic or a nonmember. Members also share substantial portions of their follower networks with fellow pack members. Despite this coherence, a typical member's follower follows only a small fraction of other members in the pack, indicating limited audience coverage. These results suggest that starter packs capture coherent topical and social groups while leaving opportunities for further account discovery.
More broadly, results show human-curation can be useful to capture relevant communities and connections, and platforms could integrate human-curation into recommendation systems as a complement to algorithmic systems to expand opportunities for discovery.

\end{abstract}


\section{Introduction}
Social media platforms are powerful tools for community building, enabling people to connect based on shared interests and backgrounds. However, these communities are tied to the platform. Changes in platform policies or ownership may cause users to move to other platforms or de-platforming through suspension \cite{alizadeh2022content,papasavva2024waiting,ng2025journalists,ali2021understanding}. 
A recent example is the 2022 Twitter/X acquisition \cite{radivojevic2025user,jeong2024exploring}. After the acquisition, many users moved to alternatives such as Mastodon, and later Bluesky and Threads. 
A prominent example of such migration happened for scientific community on these social network platforms \cite{kupferschmidt2024researchers,quelle2025academics}.
However, arriving on a new platform means rebuilding connections and a community to engage with an exchange information. This creates a need for social media platforms to offer features that help new users discover people to follow.

Bluesky, a decentralized microblogging platform launched in 2023, introduced starter packs \cite{bluesky_starterpack_launch}. These starter packs consist of human-curated lists of accounts and feeds designed to help new users quickly find people worth following \cite{balduf_bootstrapping_2025}. Anyone can create and share packs, and users can follow the recommended accounts together. These feature allows both newcomers and existing users to step into a community or topic area. At the same time, the same mechanism can also affect who is more visible to new users, if they are more frequently considered by starter pack curators.

How curators select users and exclude some can impact visibility and influence of accounts as previous research by Balduf et al. demonstrated \cite{balduf_bootstrapping_2025}. Therefore, curators of starter packs influence not only which accounts are grouped together but also who receives additional exposure. It is also known that visibility is not equally distributed across demographic groups \cite{nilizadeh2016glassceiling, chakraborty2017trends} and skews towards the popular accounts \cite{salganik2006experimental, barabasi1999emergence}. These network effects also influence the recommendation algorithms and create biases towards more connected accounts \cite{klimashevskaia2024survey} and personalization is one of the approaches to mitigate that effect \cite{abdollahpouri2017controlling,abdollahpouri2019popularity,zhu2021popularity}.

Starter packs are positioned between algorithmic recommendation and organic network formation. They do not optimize for engagement; however, curators could be biased towards those external signals. 
Starter packs are also not communities which emerge from the follow graph over time. 
A starter pack curator makes an explicit, deliberate choice: these people belong together. With that choice, they decide who should be seen by a newcomer, and who shouldn't. 
Human curated lists can identify meaningful topical groups \cite{bhattacharya2014deep, bhattacharya2014inferring}. Their composition may also reflect curators' exiting social ties and demographic homophily, given the tendency for social relationships to connect similar people \cite{mcpherson2001birds}. 
As a result, understanding starter packs require examining both the relationships that brings the members together, and the selections patterns of the curators. Do packs identify accounts that belong together? Do they reflect their curator's statements about the pack itself? How does their membership correspond to existing follow relationships?

In this paper, we study 2,718 active starter packs covering 144,987 unique accounts. We combine account posting histories, pack descriptions, follower networks, and inferred demographic attributes to examine three research questions:

\begin{itemize}
    \item \textbf{RQ1:} Are starter-pack members more aligned with their own pack than comparable accounts from the same broader topic?
    \item \textbf{RQ2:} Do pack curators resemble the accounts they include?
    \item \textbf{RQ3:} How are starter-pack members connected through shared audiences and follow relationships?
\end{itemize}

\begin{figure}[t]
  \centering
    \begin{tikzpicture}[
      objectbox/.style={
        draw=black!80,
        fill=black!3,
        line width=0.55pt,
        rounded corners=1pt,
        text width=7.55cm,
        inner sep=2.2mm,
        align=left,
        font=\scriptsize
      },
      primaryedge/.style={
        -{Stealth[length=2.0mm,width=1.5mm]},
        draw=black,
        line width=0.95pt
      },
      secondaryedge/.style={
        -{Stealth[length=1.8mm,width=1.3mm]},
        draw=black!55,
        line width=0.65pt,
        dashed
      },
      edgelabel/.style={
        fill=white,
        inner xsep=2.5pt,
        inner ysep=1.2pt,
        font=\scriptsize
      },
      notebox/.style={
          draw=black!45,
          fill=white,
          line width=0.4pt,
          rounded corners=1pt,
          text width=7.55cm,
          inner sep=2mm,
          align=left,
          font=\scriptsize,
          dash pattern=on 1.6pt off 1.4pt
        },
      node distance=14mm
    ]
    
    \def\field#1{%
      \hangindent=1.15em
      \hangafter=1
      \noindent
      \makebox[1.15em][l]{\textbullet}#1\par
    }
    
    \node[objectbox] (pack) {%
      \parbox{7.55cm}{%
        \raggedright
        {\footnotesize\textbf{Starter pack}\hfill
         \textbf{\(n=2{,}718\)}}\par
        \vspace{0.6mm}\hrule\vspace{0.9mm}
        \textit{Human-curated list of accounts and feeds that a new user can follow.}\par
        \vspace{0.8mm}
        \field{Name and description}
        \field{Topic label (one of 17 LLM-assigned categories)}
        \field{Join count (cumulative users joining via the pack)}
        \field{Member list}
        \field{Curator account (also a member in \(2{,}713\) of \(2{,}718\) packs)}
      }%
    };
    
    \node[objectbox, below=of pack] (member) {%
      \parbox{7.55cm}{%
        \raggedright
        {\footnotesize\textbf{Member account}\hfill
         \textbf{\(n=144{,}987\)}}\par
        \vspace{0.6mm}\hrule\vspace{0.9mm}
        \textit{Account appearing in at least one starter pack.}\par
        \vspace{0.8mm}
        \field{Posts: originals, reposts, and quote posts (replies collected but not embedded)}
        \field{Follower count}
        \field{Follower and follow edges}
        \field{Inferred gender, age bracket, and organization status}
        \field{Inferred race}
        \field{Text embedding vector constructed from the account's posts}
      }%
    };
    
    \node[notebox, below=7mm of member] (corpus) {%
      \parbox{7.55cm}{%
        \raggedright
        {\footnotesize\textbf{Post corpus}}\par
        \vspace{0.6mm}\hrule\vspace{0.9mm}
        \begin{tabular}{@{}r@{\hspace{1.6mm}}l@{}}
          \(218{,}557{,}938\) & collected rows (one feed item per member)\\
          \(215{,}852{,}145\) & distinct (member, post) pairs\\
          \(165{,}760{,}715\) & distinct posts\\
        \end{tabular}\par
        \vspace{0.8mm}
        Rows exceed pairs by \(2{,}705{,}793\) refetch duplicates; pairs exceed
        distinct posts because a reposted post is counted once per reposting member.
      }%
    };
    \draw[secondaryedge] (member.south) -- (corpus.north);
    
    \draw[primaryedge]
      (pack.south) --
      node[edgelabel, font=\scriptsize\bfseries]
        {pack--member edges: \(n=192{,}491\)}
      (member.north);
    
    \end{tikzpicture}
    \caption{Data collection and processing summary.}
    \label{fig:tikzchart}
\end{figure}
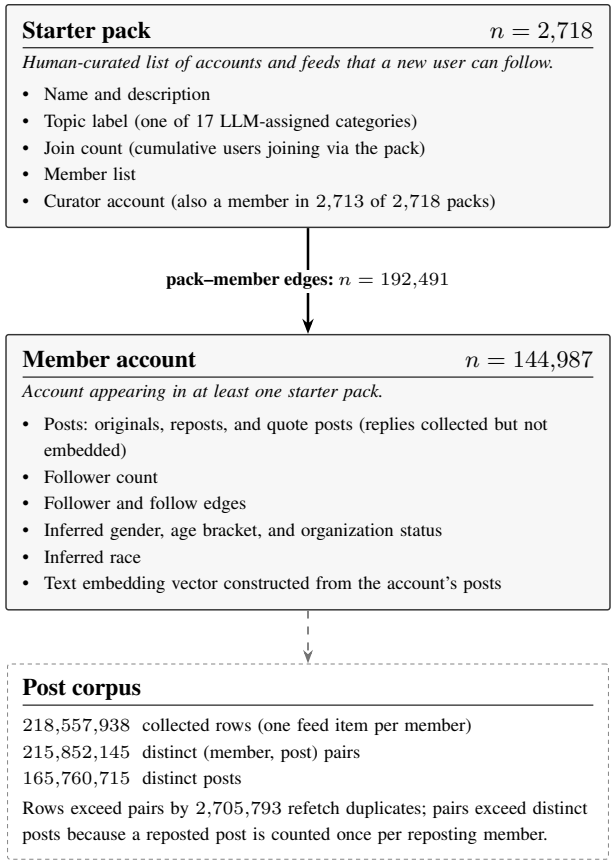

To address these questions, we collected comprehensive collection of starter packs and accounts associated with them as summarized in Fig.\ref{fig:tikzchart}. Our data include approximately 165.8 million distinct posts collected over two rounds between December 2025 and May 2026. We categorize packs using Gemini 3.5 Flash, and embed account posts and pack descriptions using \texttt{EmbeddingGemma300M} in order to measure semantic similarity. We examine demographic resemblance using attributes inferred with M3Inference \cite{wang2019m3} and ethnicolr \cite{chintalapati2023predictingraceethnicitysequence}. We finally utilize follow networks to capture shared followers, proportions of pack members reached, and reciprocity.

Our findings suggest that members align more closely with their own pack than accounts that belong the same broader topic, and accounts within the same pack are semantically more similar than pairs of random accounts from the same broader topic. Members also resemble with their curators semantically, while demographic resemblance varies across attributes and topics. In the follow network, we show that members share audiences and have higher reciprocity towards fellow pack members and curators of the packs that they belong in. 

These findings show that starter packs capture topical and social structure within broader topic areas, suggesting their potential to support discovery beyond initial onboarding and direct link sharing by curators. Because accounts following one or more members follow only a small fraction of the pack on average, Bluesky could consider suggesting a relevant starter pack to users who already follow one or more of its accounts. Such an approach may help users to discover related accounts beyond their immediate follow network while preserving the curator's framing. 

\section{Related Work}
\subsection{Platform migration and the growth of Bluesky}
The 2022 acquisition of Twitter triggered waves of user departure to alternative microblogging platforms. Research on migration to Mastodon shows that these movements were shaped by existing social connections, emphasizing the importance of network dynamics in order to understand platform migrations \cite{he2023flocking}. Therefore, migration also means that a user has to establish some connection with the new platform.

Bluesky launched private beta in 2023 and opened registration to the public in February 2024 \cite{bluesky_beta_2023, bluesky_public_launch_2024}. Its user base grew from three million at the public launch to over 25.9 million by December 2024 \cite{bluesky_public_launch_2024, bluesky_year_review_2024}. This growth provides the opportunity to study how newcomers to a new platform discover new accounts and build their new network \cite{seckin2025rise}. Particularly, migration of academic community has been studied extensively \cite{kupferschmidt2024researchers,quelle2025academics}.

\subsection{Cold start and starter packs}
New users of a new platform present a major problem: they have little to no connections on the platform they have landed in, which limits the platform's ability to personalize the experience. Studying Pinterest and Last.fm, \cite{zhong2014social} found that connections copied from Facebook supported social interaction and helped the onboarding process. However, active users formed new connections in the platform eventually. Therefore, even if it helped, copying a graph does not replace the organic discovery within the destination platform.

Bluesky introduced starter packs in June 2024 as a way of smoothing the onboarding process. They are shareable collections that recommend accounts and custom feeds, allowing users to follow a set of accounts and feeds together \cite{bluesky_starterpack_launch}. Starter packs can be created by anyone without and algorithmic assistance, therefore they are a form of non-algorithmic recommendation.

\cite{balduf_bootstrapping_2025} examined 335,416 starter packs and more than 25 million users. They found that starter packs were responsible up to 43\% of daily follow operations at their peak, and used propensity score matching to quantify the impact. They identified that being added to a starter pack bring benefits such as an increase in the follower count. Which results in those users being more active in the platform and might reinforce a rich get richer effect. 

\subsection{Human curation and social groups}
Human-curated account lists have a history before Bluesky. Bhattacharya et al. \cite{bhattacharya2014deep, bhattacharya2014inferring} used Twitter List metadata to identify topical experts and groups, and to infer users' interests from the experts they follow, demonstrating that curated lists can reveal meaningful topical structure. \cite{sharma2012inferring} used lists to infer a user’s expertise and subsequently, Ke et al. \cite{ke2017scientists} identified scientists and inferred their disciplinary affiliations from community-generated list labels. \cite{benabdelkrim2020finding} constructed a network of Twitter users from shared list memberships and proposed a method for identifying local user communities and their common interests.

Google+ circle sharing allowed users to recommend groups of users to others. Fang et al \cite{Fang2012Look} examined circles using member popularity, reciprocity of follows and density of mutual connections among members. They distinguished community circles, characterized with high owner reciprocity and relatively low member popularity, and celebrity circles, characterized with lower owner reciprocity and higher popularity. After circle-sharing events, community circles experienced faster growth in mutual connections and larger increases in connections involving users with fewer existing connections. Brauer and Schmidt \cite{Brauer2014Circles} found that circles exhibited distinct network structure while preserving connections outside the group. These findings suggest that curated collections can capture social relationships while maintaining connections to the wider network.

Network gatekeeping theory provides a framework for examining who controls information selection and circulation \cite{barzilainahon2008toward}. In starter packs, curators can be treated as gatekeepers controlling which users receive exposure to newcomers. Although exclusion does not prevent discovery elsewhere on the platform, curators' choices shape how a topic or community is presented to newcomers.

These choices may also reflect homophily, the tendency for social connections to occur between similar people \cite{mcpherson2001birds}. Similarity can involve common interests or demographic characteristics, and can arise from both individual preferences and social settings that make such contact more likely. In starter packs, resemblance between curators and members may reflect topical specialization, shared interests, existing social ties, or demographic similarity. This motivates examining semantic and demographic resemblance separately.

\subsection{Bias and fairness}
Previous research shows that visibility varies systematically across demographic groups. \cite{nilizadeh2016glassceiling} examined followers, list memberships and retweets on Twitter, and found that perceived gender had different associations with visibility at different levels of popularity, stating a disadvantage for users perceived as female among the most visible accounts. Similarly, \cite{chakraborty2017trends} found that the users promoting trends on Twitter were often unrepresentative of the platform population, and some groups are systematically underrepresented.

Recommendation algorithms can reinforce these inequalities. \cite{stoica2018glassceiling} showed that algorithms can amplify demographic underrepresentation. Research on fairness-aware recommendation keeps growing in order to tackle these issues, with studies distinguishing multiple fairness objectives and evaluation approaches \cite{jin_survey_2023, Wang_Ma_Zhang_Liu_Ma_2023, Pessach_Shmueli_2022}. Although starter pack membership is a result of human selection, differences in who receives exposure remain relevant.

Popularity is another factor in unequal visibility. Preferential attachment describes how new connections are more likely to attach to already well-connected nodes \cite{barabasi1999emergence}. In starter packs, inclusion itself can increase follower counts and activity \cite{balduf_bootstrapping_2025}, complicating comparisons between members and nonmembers.  

\section{Data and Methods}
\label{sec:data-method}

\begin{figure}[t]
    \centering
    \includegraphics[
        width=\columnwidth,
        keepaspectratio
    ]{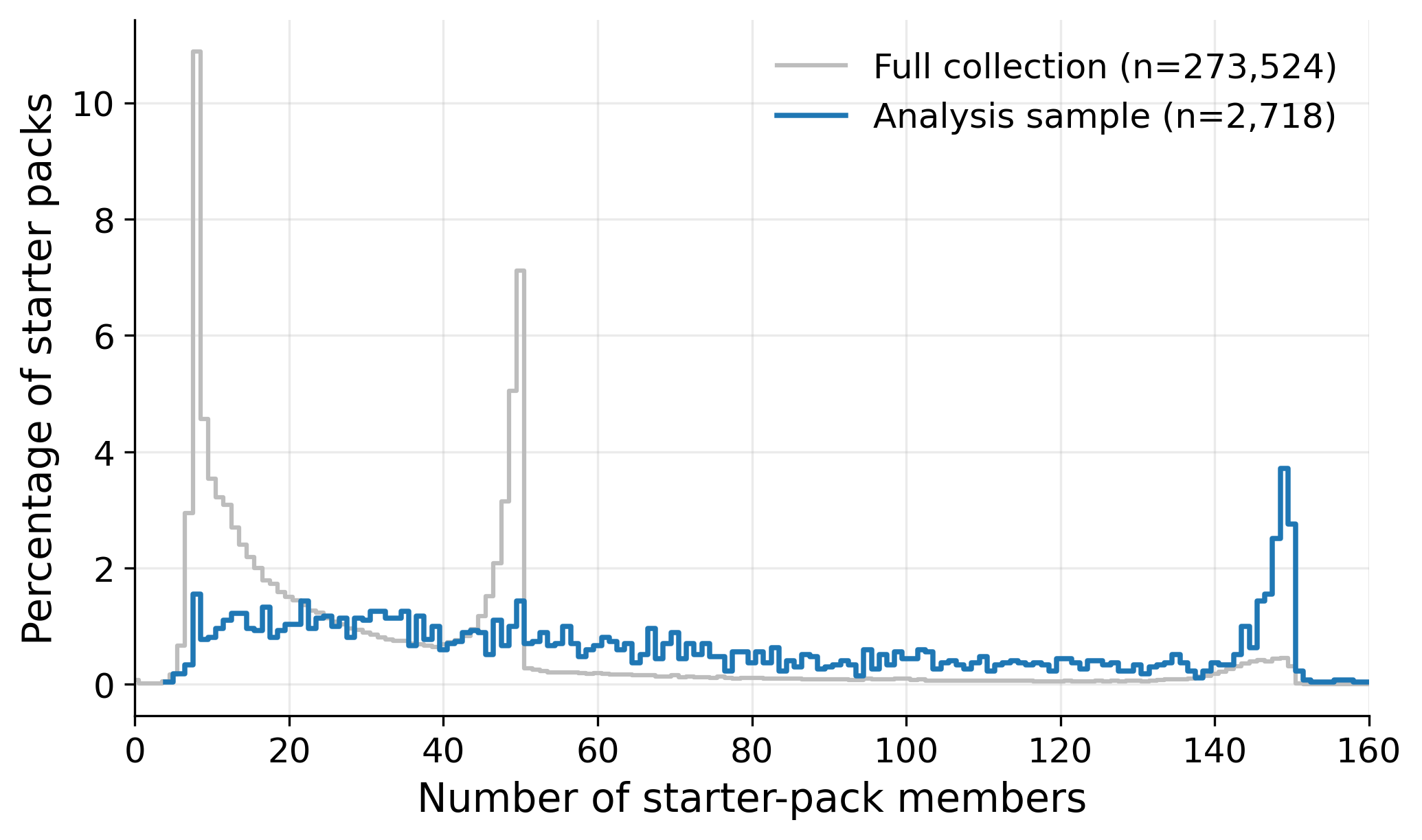}
    \caption{more than 150 members in analysis sample: 18 of 2,718 packs: 0.662\%. more than 150 members in full collection: 265 of 273,524 packs: 0.097\%}
    \label{fig:member_distribution}
\end{figure}

\subsection{Data collection}
Starter pack URIs were identified from two public aggregator websites, Bluesky Directory\footnote{\url{https://blueskydirectory.com/starter-packs}} and Bluesky Starter Packs \footnote{\url{https://blueskystarterpack.com}}, yielding 310,435 unique URIs. 
Pack metadata and member lists were collected via the Bluesky AT Protocol API for collecting starter pack details\footnote{\url{app.bsky.graph.getStarterPack}}and list of members \footnote{\url{app.bsky.graph.getList}}, resulting in 273,524 successfully fetched packs. 
We use the API field \texttt{joined\_all\_time}, the cumulative count of users who joined a starter pack, as our measure of pack popularity; we refer to this quantity as the \emph{follower count} throughout. 
We restrict analysis to packs with at least 5 followers, resulting with 2718 packs or approximately 0.99\% of the successfully fetched packs, as packs below this threshold have not been meaningfully adopted by platform users.

After the filtering, the analysis dataset contains information for these 2,718 packs and their 192,491 pack-member edges corresponding to 144,987 unique accounts.
Curators appear in their own pack in 2,713 of 2,718 packs. Posts were collected in two batches: the first retrieved full available author histories between December 2025 and February 2026, and the second retrieved posts dated after 30 November 2025 in May 2026. The overlap was deduplicated by post URI, retaining the first-batch copy when a post appears in both. The corpus contains nearly 218.6M rows initially and 165.8M distinct post URIs after processing.

Pack size ranges from 4 to 300 members, with a median of 59 as shown in Fig.\ref{fig:member_distribution}. 
Although Bluesky documentation describes starter packs as containing up to 150 recommended accounts, 18 of the 2718 packs in our sample contain more than 150 retrieved members. Reconstructed member counts may also be lower than a pack's reported \texttt{listItemCount}, because accounts deleted or suspended after curation are not returned by the API.

We also collect the social network among members by using API endpoints for friends\footnote{\url{app.bsky.graph.getFollows}} and followers\footnote{\url{app.bsky.graph.getFollowers}} over all 144,987 accounts in April 2026. Records were returned for 142,972 of them, giving $\approx103M$ directed follow edges.

\subsection{Assignment of pack communities}

Starter packs are organized around topics or themes introduced by the curator. They define a title and description for the pack to inform members and the followers. To study different topics, we grouped them based on the textual information available for the packs. 

BERTopic was used to produce an initial set of topics; however, 69\% of packs fell into four clusters whose top keywords were generic terms such as ``and,'' ``the,'' ``pack,'' and ``starter,'' and it produced no counterpart to several categories that are common in the data, including Politics \& Journalism, Medicine, and LGBTQ+. We therefore used it only to seed the labels, which we expanded the into a final set of 17 labels. Each pack was then assigned one label from this fixed set by Gemini 3.5 Flash, prompted with the pack's title and description. To validate the assignments, one of the authors labeled a sample of 200 packs and assign one of the 17 categories, giving Cohen's $\kappa = 0.912$ agreement.

\begin{figure}[t]
    \centering
    \includegraphics[
        width=\columnwidth,
        keepaspectratio
    ]{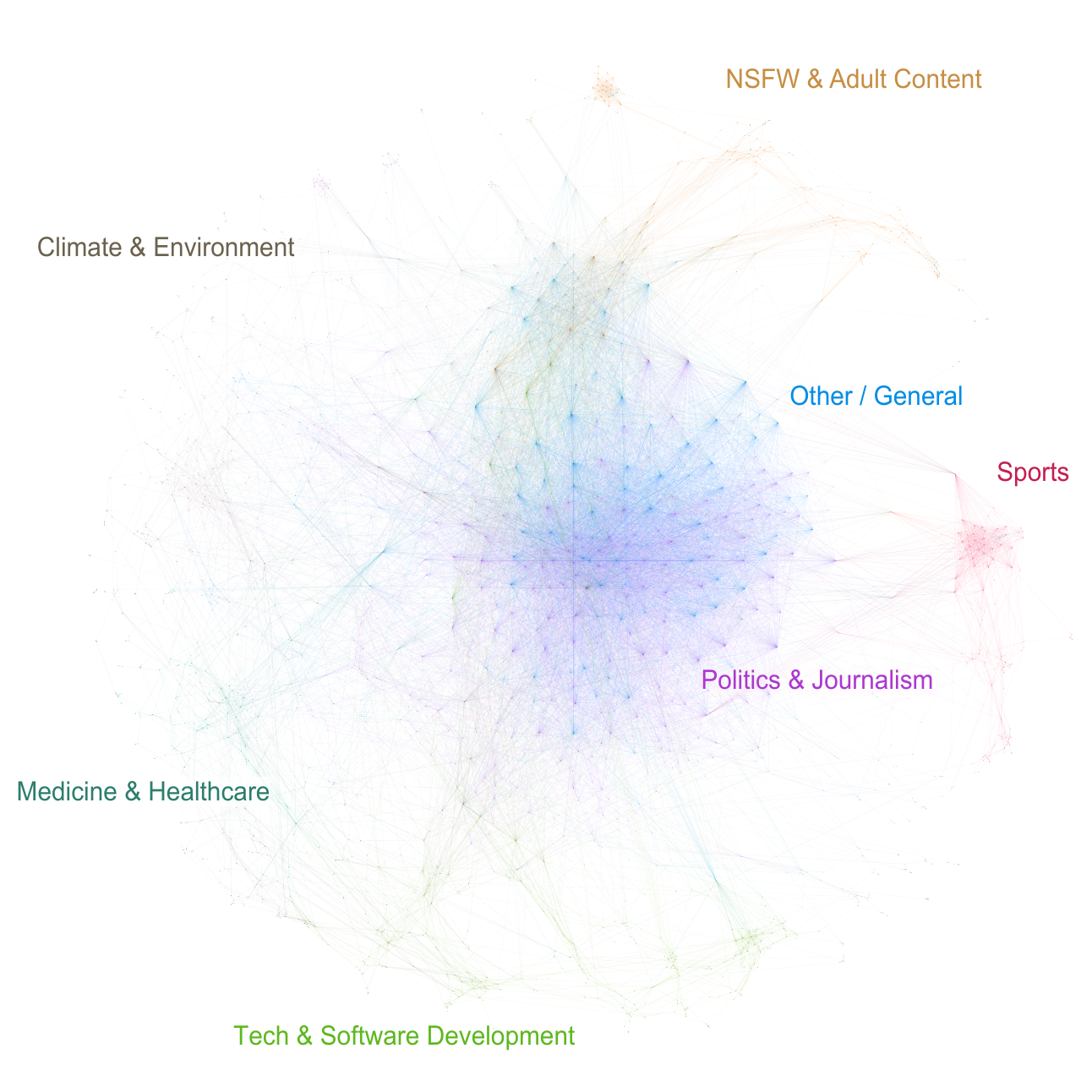}
    \caption{Largest connected component of the pack–pack overlap graph (2,063 of 2,718 packs). Nodes are starter packs and their colored represents their topic labels. Edges connect node pairs with member overlap $\geq 0.05$. 
    \texttt{ForceAtlas2} algorithm is used for layout and labels are shown for the prominent categories only, which correspond to $\approx70\%$ of the packs on the network.
    }
    \label{fig:network}
\end{figure}

To visualize the structure of the pack collection, we construct a pack overlap network in Figure \ref{fig:network}. In this network, nodes represents starter packs and edges connect two packs quantified by the overlap coefficient of their member sets, $\frac{|M_i \cap M_j|}{\min(|M_i|, |M_j|)}$, where $M_i$ represents set of members for starter pack $i$. We set a threshold of 0.05 to better display the structure of the network.

\subsection{Embeddings for members and packs}
To obtain a dense representations for starter packs and their members, we use Google's EmbeddingGemma model\footnote{\url{https://huggingface.co/google/embeddinggemma-300m}} specifically prompted to produce embeddings for clustering tasks, yielding a 768-dimensional and L2 normalized vectors. Textual metadata of starter pack title and description used to embed packs, while accounts are represented with their posts' embeddings. 

Bluesky feeds return four kinds of items: \textit{reposts}, which relay another account's post without commentary; \textit{quote posts}, which report with additional content created by the account; \textit{replies}, which attach to a parent post within a thread; and standalone posts, which we refer to as \textit{originals}. A post can technically be both a reply and a quote; we classify these posts as quotes. 
We embed originals, reposts, and quotes, while excluding replies. Replies are anchored to their thread context, so their content often reflects concerns raised by the parent author rather than topics the replying account engages with on its own initiative. For reposts and quotes, we attribute the referenced post's text to the account that shared it, on the grounds that amplification is itself part of what an account presents to its followers.

Images and other media are excluded from the embedding pipeline. A post carrying an image contributes only with its text, and a post with no text (3.4\% of all posts) does not contribute. Posts with fewer than five words (7.8\% of all posts) are also excluded, as very short text gives an unstable vector. 
Bluesky limits posts to 300 graphemes, so the 300-character truncation we apply rarely binds, affecting 
only 0.2\% of the posts.
A quote contributes two components, each truncated separately, so a joined quote input can reach 601 characters. Inputs are capped at 256 tokens for originals and reposts and 512 for quotes to capture content for all text.
At the end 88.8\% of all posts produce a vector for the analysis.

An account's vector is the arithmetic mean of its eligible post vectors. As pack owners are members of their own packs, they do not receive a different treatment. 
Of the 144,987 accounts in the analysis set, 133,786 (92.3\%) have a post vector; the remaining 11,201 have no post surviving the inclusion criteria above. Pack centroids are therefore means over the covered members: the median pack has vectors for 96.6\% of its members.

We define a pack's text vector as the embedding of its name and description concatenated. Of the 2,718 packs, 282 (10.4\%) have an empty description, so their text vector is the name alone. A pack's member centroid is the arithmetic mean of the vectors of its members. Vectors are normalized when cosine similarity is computed. Because account vectors are not normalized before averaging, accounts with larger vector norms contribute more strongly to the direction of the pack centroid.

\begin{figure}[t]
    \centering
    \includegraphics[
        width=\columnwidth,
        keepaspectratio
    ]{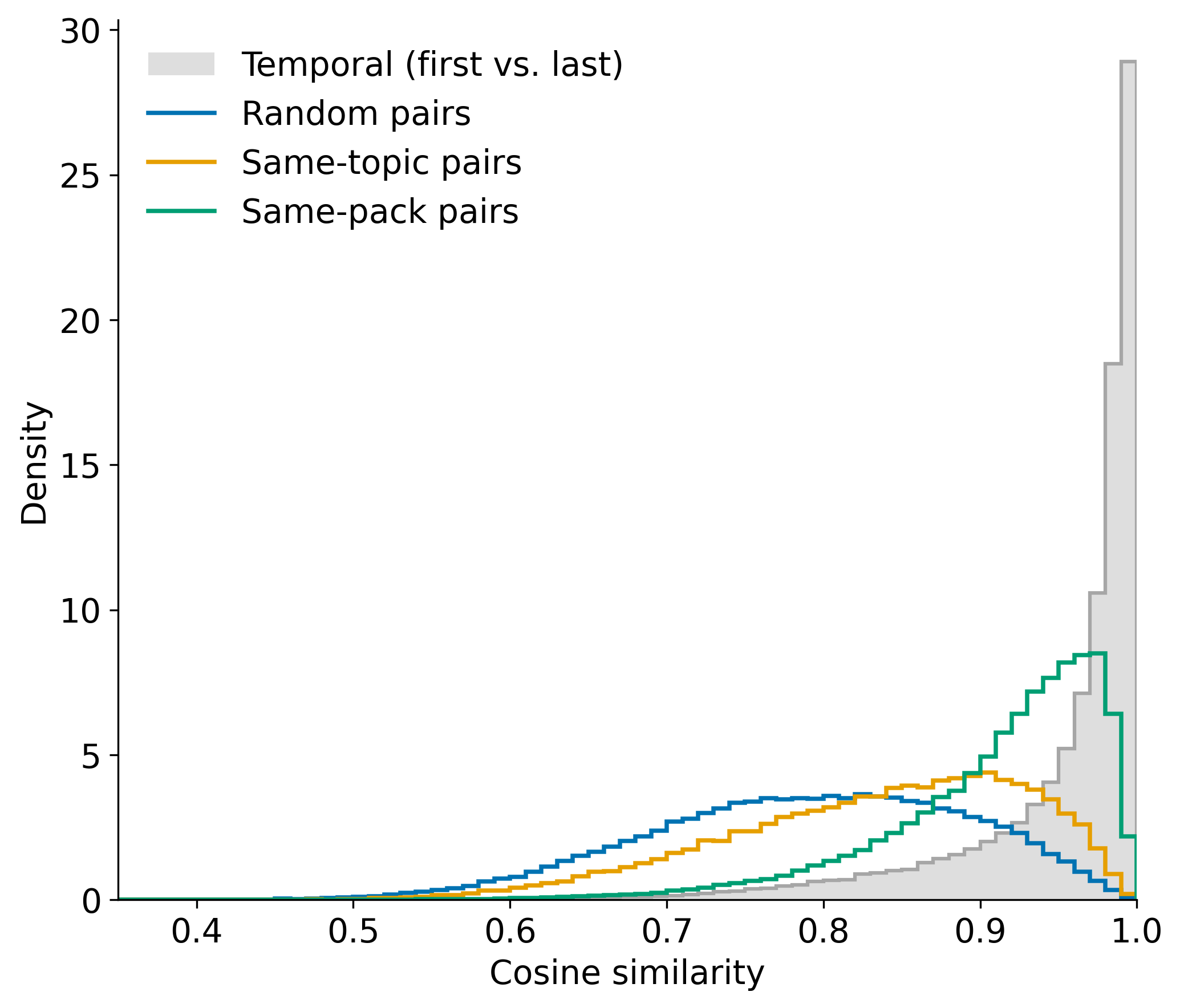}
    \caption{Cosine similarity of member pairs and accounts with
themselves over time. Three lines show the densities of the similarity between two different accounts, whereas filled histogram is the similarity of a single account's first 25\% of posts to its last 25\%.}
    \label{fig:cosine_distributions}
\end{figure}


We run two validity checks on the account vectors.
The first check compares cosine similarity across three tiers of account pairs: random pairs, pairs sharing a topic, and pairs sharing a starter pack as shown in Figure~\ref{fig:cosine_distributions}. Mean similarity rises from 0.789 for random pairs to 0.832 within a topic and 0.909 within a pack; the corresponding medians are 0.794, 0.846, and 0.928. The ordering holds within every topic individually, so it is not a result of a few unusually cohesive communities. 
We also note that pairwise similarity patterns hold when considering only original posts, indicating that reposting and quoting behavior does not drive these similarities.
Pack co-membership also carries information beyond the topic label. We see that among same-topic pairs, those share a pack average 0.907, meanwhile those that do not average 0.831.
We note that absolute similarity values are not interpretable on their own as even random account pairs average 0.789. This is a consequence of the model's output occupying a narrow region of the embedding space, because of a geometric phenomenon called anisotropy. As a result, only the differences and ordering carry meaning, and only against a fixed reference. Thus, we never compare cosines to a post centroid and cosines to a pack description, as they have different scales.


The second check investigates temporal stability by comparing account vectors constructed from the first and last quarters of each account's chronologically ordered post history. We require at least 20 posts in each window before text filtering, and cap each window at 500 posts.
After the filters, 75,194 accounts (51.0\% of all starter pack members) remain. Cosine similarity between the two window centroids has a mean of 0.979 and a median of 0.985 (IQR: 0.973--0.992).

Agreement remains high across all 17 topics, with topic-level median ranging from 0.979 to 0.989. These results suggest that analyzed accounts' earlier and later posts are broadly similar. Although they do not establish unchanged posting behavior or extend to accounts excluded by the eligibility threshold. Agreement also increases with embedded-post count with the median similarity rising from 0.971 in the lowest post count group to 0.992 in the highest. Therefore, posting volume should be taken into consideration when interpreting cosine-based results.


This analysis suggests that we can effectively learn representations for posts. Furthermore, account and pack embeddings derived from post content and metadata reliably differentiate accounts by topic.

\section{Results}

To address our research questions, we collected starter pack details, post and account level information. To systematically discuss the results, we follow the following notation: A pack $p$ in topic $t$ contains set of members denoted as $M_p$. The curator of the pack is $c_p$ and $U_t$ represents the set of unique accounts appearing in any analyzed pack within the topic $t$.
Using the pack and account embeddings, we defined alignment measures:
\begin{align}
a^{\mathrm{cent}}_{u,p}
&= \cos(\mathbf{e}_u,\boldsymbol{\mu}_{p,-u}),\\
a^{\mathrm{desc}}_{u,p}
&= \cos(\mathbf{e}_u,\mathbf{d}_p),
\end{align}

where $\mathbf{e}_u$ is account $u$'s embedding, and $\mathbf{d}_p$ is the embedding of the pack's title and description. The member centroid $\boldsymbol{\mu}_{p,-u}$ excludes $u$ when $u \in M_p$ to avoid self-inclusion. Otherwise, the full pack centroid is used. The curator $c_p$ is excluded from $a^{\mathrm{desc}}_{u,p}$ for their own packs because they authored its title and description. We refer to these measures as member alignment and description alignment, and analyze them separately.

In discriminative analysis, the Area Under the ROC Curve (AUC) quantifies how well a model's predicted scores rank relevant items ahead of non-relevant ones across all possible decision thresholds, which is equivalent to the Wilcoxon–Mann–Whitney statistic. AUC represents the exact probability that a randomly selected positive instance will receive a higher score or rank than a randomly selected negative instance. 
We use this measure to quantify how well users with particular property or measurement differentiate from others.
An AUC of $0.5$ indicates no ranking tendency while values about $0.5$ indicate higher scores for members. Topic-level AUC scores are unweighted means of the pack-level values.

\subsection{RQ1: Are starter-pack members more aligned with their own pack than comparable accounts from the same broader topic?}
For each pack $p$ in topic $t$, we compare member and description alignment between $M_p$ and $U_t \setminus M_p$, applying the curator exclusions defined above.

\begin{table*}[!th]
\centering
\small
\caption{Topic-level alignment and curator-similarity results. \emph{Centroid}, \emph{Description}, and \emph{Semantic} are AUCs (0.5 = no signal, higher = members rank above non-members). \emph{Age}, \emph{Gender}, and \emph{Race} are skews in percentage points of differences between the curator-matching proportions between members and the same-topic reference. All values are unweighted means over the packs in a topic. Skews are significant at $p<0.05$ by pack-level permutation test unless marked $^{\dagger}$.}
\label{tab:topic_alignment_owner}
\begin{tabular}{lcccccc}
\toprule
& \multicolumn{2}{c}{RQ1: Pack alignment}
& \multicolumn{4}{c}{RQ2: Curator resemblance} \\
\cmidrule(lr){2-3}\cmidrule(lr){4-7}
Topic & Centroid & Description & Semantic & Age & Gender & Race \\
\midrule
Academia \& Research          & 0.837 & 0.818 & 0.802 & $+4.54$ & $+2.87$ & $+3.68$ \\
Artists (Visual, SFW)         & 0.842 & 0.800 & 0.798 & $+11.32$ & $+12.38$ & $+3.27$ \\
Authors \& Literary           & 0.784 & 0.726 & 0.740 & $+5.22$ & $+9.39$ & $+0.96^{\dagger}$ \\
Biology \& Life Sciences      & 0.842 & 0.867 & 0.835 & $+2.84$ & $+4.27$ & $+3.54$ \\
Climate \& Environment        & 0.836 & 0.830 & 0.801 & $+2.49$ & $+4.29$ & $+3.56$ \\
Educators (K-12)              & 0.847 & 0.791 & 0.840 & $+1.48^{\dagger}$ & $+6.09$ & $+2.94$ \\
Fan Communities               & 0.845 & 0.809 & 0.793 & $+1.42^{\dagger}$ & $+7.53$ & $-0.69^{\dagger}$ \\
Foreign Language Communities  & 0.954 & 0.898 & 0.937 & $+9.87$ & $+10.07$ & $+17.80$ \\
Gaming \& Game Development    & 0.781 & 0.778 & 0.754 & $+5.29$ & $+3.13$ & $+2.08^{\dagger}$ \\
LGBTQ+ Community              & 0.823 & 0.725 & 0.823 & $+4.18^{\dagger}$ & $+5.87^{\dagger}$ & $-8.17$ \\
Medicine \& Healthcare        & 0.816 & 0.841 & 0.786 & $+2.39$ & $+6.96$ & $+1.76^{\dagger}$ \\
Music                         & 0.754 & 0.730 & 0.725 & $-2.80^{\dagger}$ & $+6.89$ & $+3.13^{\dagger}$ \\
NSFW / Adult Content          & 0.836 & 0.774 & 0.818 & $+4.32$ & $+14.62$ & $-1.63^{\dagger}$ \\
Other / General               & 0.797 & 0.737 & 0.762 & $+6.77$ & $+12.26$ & $+1.28^{\dagger}$ \\
Politics \& Journalism        & 0.882 & 0.850 & 0.853 & $+5.23$ & $+4.57$ & $+1.69$ \\
Sports                        & 0.903 & 0.864 & 0.781 & $+0.61^{\dagger}$ & $+4.77$ & $+0.35^{\dagger}$ \\
Tech \& Software Development  & 0.819 & 0.813 & 0.769 & $+4.72$ & $+3.90$ & $+1.37$ \\
\bottomrule
\end{tabular}
\end{table*}

As shown in Table~\ref{tab:topic_alignment_owner}, members show substantially higher alignment with their own pack than same topic comparison accounts. This pattern holds across all 17 topics with topic-level mean AUCs ranging from $0.754$ to $0.954$ for member alignment and from $0.725$ to $0.898$ for description alignment. 

Together with the pairwise similarity comparisons reported in Figure~\ref{fig:cosine_distributions}, these results suggest that starter packs capture topical specificity beyond the broader topic label. Members align both with the pack's member composition and with its curator-authored title and description.  

When topics with high and low alignment are investigated, Foreign Language (0.954), Sports (0.903), and Politics \& Journalism (0.882) rank at the top, while Gaming (0.781) and Music (0.754) have lower alignment with the broader topic, indicating that users in these categories may have more diverse interests. Pack alignments are also stronger with member centroids than with pack descriptions, with the sole exception of Medicine \& Healthcare topics. This suggests that members better represent the packs for platform users, while highlighting curators' ability to assemble coherent accounts. 

A broader implication of these findings is that platforms like Bluesky could recommend users from similar packs or algorithmically rank content from topically adjacent packs higher.

\subsection{RQ2: Do pack curators resemble the accounts they include?}
We examine curator--member resemblance in semantic and demographic dimensions. Semantic resemblance between curator $c_p$ and account $u$ is measured as
\[
s_{p,u} = \cos(\mathbf{e}_{c_p},\mathbf{e}_u),
\]
For each pack, we compare these scores between members in $M_p \setminus \{c_p\}$ and same-topic comparison accounts in $U_t \setminus M_p$ using AUC. Curator is excluded from from both groups to avoid self-similarity. Higher similarity in this analysis suggests that curators select members that produce or share content similar to that of the curator.

Demographic resemblance is measured separately for age, race and gender. For each demographic attribute $d$, $C_{p,d}$ is the proportion of members (excluding the curator) whose inferred label matches the curator's (see Table~\ref{tab:demographic_composition} for summary statistics). We compare this with $R_{p_d}$, the average match rate across random samples of same-topic comparison accounts. To assess alignment between curators and members, we compared the observed distribution of demographic groups against expected baseline values. We define demographic skew as
\[
\Delta_{p,d}=C_{p,d}-R_{p,d}.
\]
where a positive $\Delta_{p,d}$ indicates greater resemblance to the curator than expected under the same-topic reference.

Members show greater semantic resemblance to their curators than same-topic comparison accounts across all 17 topics, with mean AUCs ranging from $0.725$ to $0.937$. Demographic resemblance is less uniform: gender skews are positive in all topics, whereas age and race show more variation in magnitude, direction and statistical significance. Foreign Language Communities has the largest positive largest skew ($+17.80$ percentage points) whereas LGBTQ+ Community exhibits a significant negative skew ($-8.17$ percentage points) in race. These findings indicate that pack members are closely aligned with their curators relative to same-topic reference both semantically and demographically. This pattern is consistent with homophily and the follow relationships we examine in RQ3 provide additional context. However, our analyses do not establish whether curator--member resemblance reflects selection preferences or relationships that preceded or developed after inclusion.

\subsection{RQ3: How cohesive and complete are the follow relationships within starter packs?}
To address this question, we examine three complementary measures: follower overlap, reciprocity, and pack fullness. These measures capture shared followers, the proportion of pack members followed by those followers, and mutual follow relationships.


\begin{figure*}[t]
    \centering
    \includegraphics[
        width=\textwidth,
        keepaspectratio
    ]{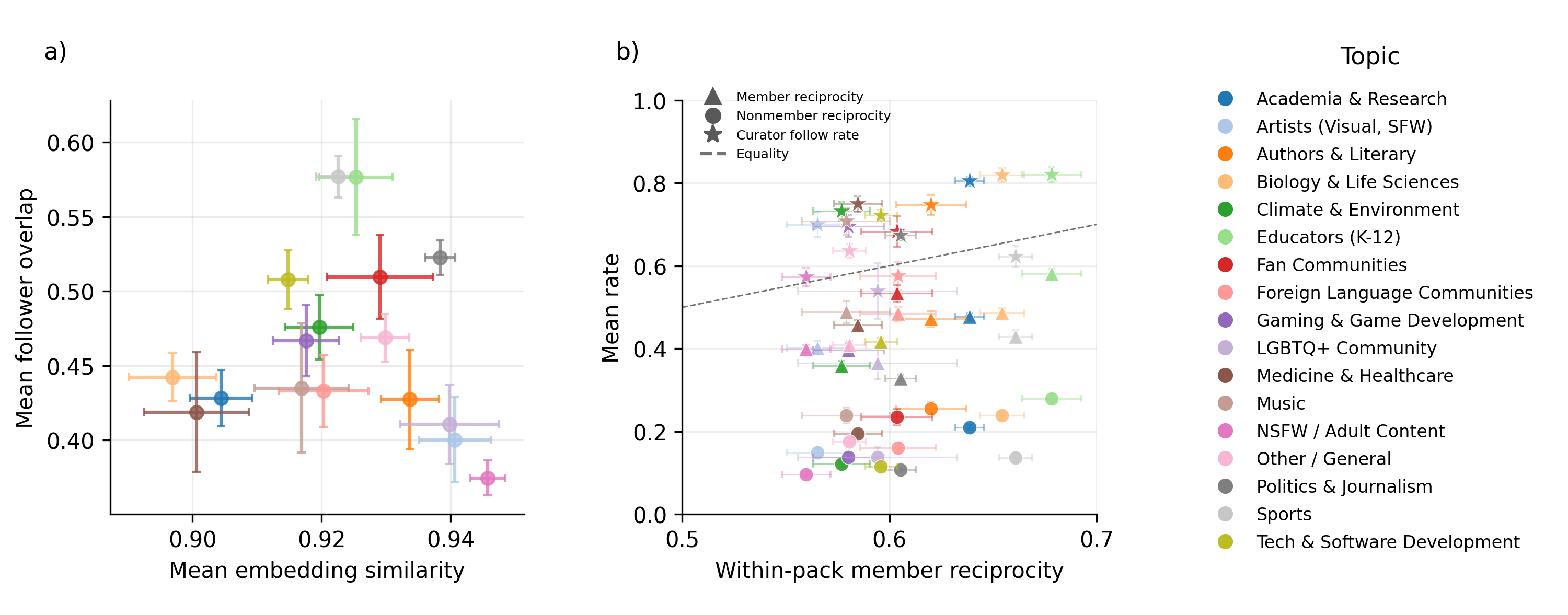}
    \caption{Audience overlap and reciprocity within starter packs. \textbf{a)} Average follower overlap coefficients and cosine similarity scores across starter packs, grouped by topic. Error bars indicate standard errors across pack. \textbf{b)} Follow relationships by topic. The x-axis shows within-pack member reciprocity. On the y-axis, circles denote nonmember reciprocity, stars denote curator follow rate, and triangles denote analysis-wide member reciprocity. Error bars show $\pm$1 standard error.}
    \label{fig:reciprocity_overlap}
\end{figure*}

For accounts $u$ and $v$, let $F_u$ and $F_v$ denote their follower sets. We measure follower overlap using the overlap coefficient.
\[
\mathrm{overlap}(u, v)
= \frac{|F_u \cap F_v|}{\min(|F_u|,|F_v|)}.
\]
We average this coefficient across all distinct member pairs within each pack, then compute unweighted means across packs within each topic.  
Pairwise cosine similarity, $\cos(\mathbf{e}_u,\mathbf{e}_v)$, is aggregated using the same procedure. 
Figure \ref{fig:reciprocity_overlap}(a) shows that the average overlap coefficient from approximately $0.35$ to about $0.60$ across topics. This indicates pack members share a considerable portion of their follower networks. However, pairwise follower overlap does not reveal how much of the pack an individual follower follows. 

As shown in Figure~\ref{fig:reciprocity_overlap}(a), topics such as NSFW / Adult Content, Artists, and LGBTQ+ Community exhibit higher content similarity alongside lower follower network overlap, suggesting more diverse user bases. A comparable level of network similarity is observed for technical topics like Medicine \& Healthcare, Academia \& Research, and Biology \& Life Sciences, though these categories display lower content similarity. Conversely, categories such as Educators (K-12) and Sports show higher levels of follower overlap, presumably because these packs primarily aggregate prominent, high-profile accounts.

Additionally, we examine how often members follow their followers back. We calculate this rate for followers belonging to any analyzed pack (member reciprocity), to the same pack (within-pack member reciprocity), and none of the analyzed packs (nonmember reciprocity). We also measure the proportion of members who follow their pack's curator (curator follow rate).

Members reciprocate follows from other starter pack members more often than from nonmembers. Within-pack members reciprocity is higher than both analysis-wide reciprocity and nonmember reciprocity (Figure~\ref{fig:reciprocity_overlap}(b)). Together with the semantic similarity results, this indicates that sharing a pack exhibits both topical coherence and stronger reciprocal connections. In most topics, curator follow rates exceed even within-pack member reciprocity.

\begin{figure}[t]
    \centering
    \includegraphics[
        width=\columnwidth,
    ]{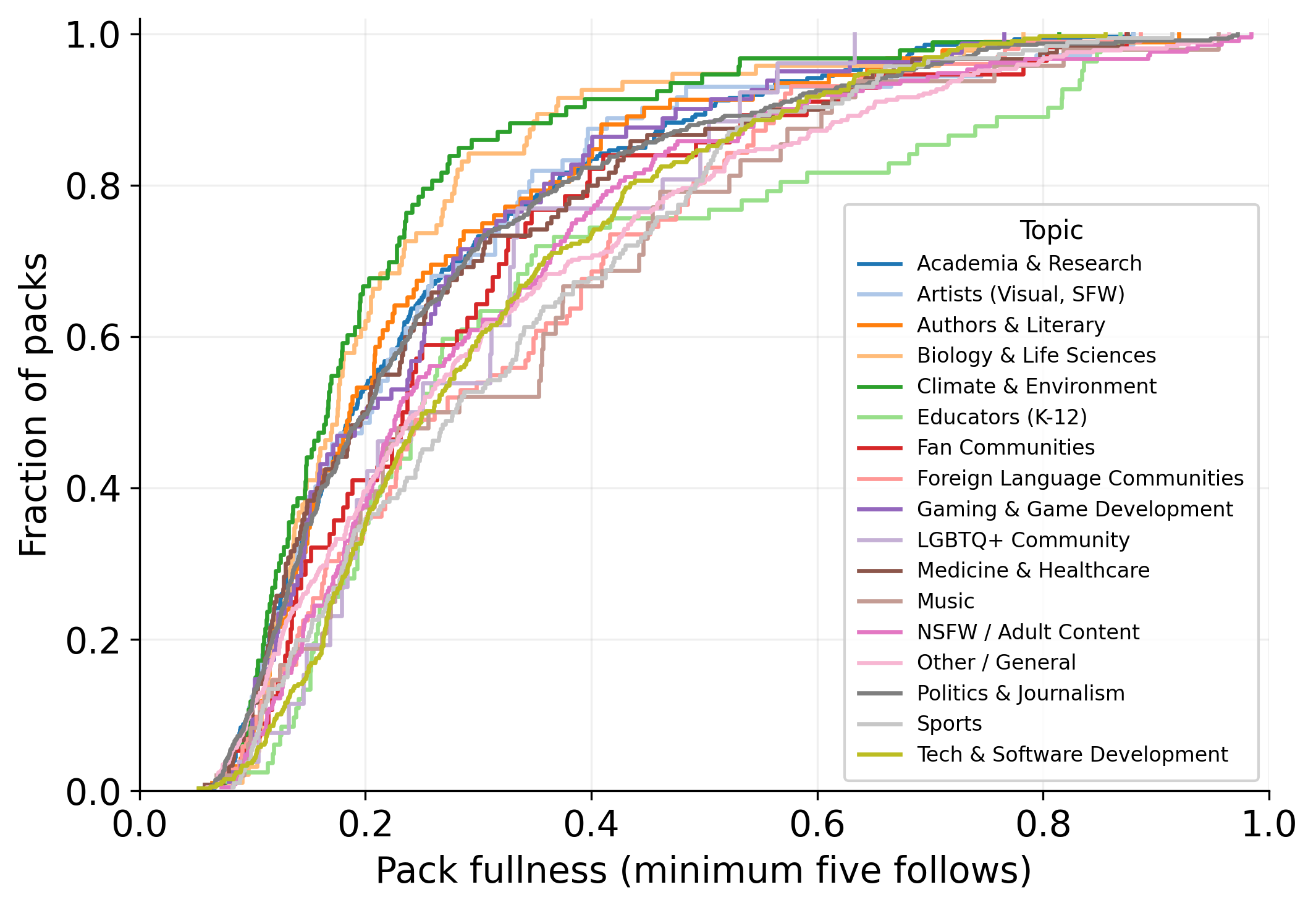}
    \caption{Distribution of pack fullness. Each line shows the CDF for one topic. Pack fullness is calculated among followers $F_p$ that follow at least 5 members of the corresponding pack. 4 of 2,718 packs were dropped from the analysis as they did not have any qualifying follower.}
    \label{fig:pack-fullness}
\end{figure}

Finally, we define pack fullness as the average proportion of members followed by accounts in the pack's collective follower set, $F_p = \bigcup_{u \in M_p} F_u$. For each account $i \in F_p$,
\[
\mathrm{fullness}_{i,p}
= \frac{|\{u \in M_p : i \in F_u\}|}{|M_p|}.
\]
Pack-level fullness is then
\[
\mathrm{fullness}_p
= \frac{1}{|F_p|}
\sum_{i \in F_p} \mathrm{fullness}_{i,p}.
\]

Pack-level fullness remains low across topics (Figure~\ref{fig:pack-fullness}). At least half of the packs have a fullness below 50\% for every topic, and the median pack across the full corpus has a fullness of $\approx0.21$. Therefore, even among accounts following at least five members, much of the pack remains outside the follow network. We note that min-five pack fullness has a strong negative correlation with pack size ($\rho = -0.849$). 
Because each qualifying follower follows at least five of the pack's $|M_p|$ members, $\mathrm{fullness}_p \geq 5/|M_p|$. This lower bound decreases with pack size, and should be considered when interpreting the negative correlation.

Shared audiences and reciprocal ties indicate that pack members are socially connected, but low fullness shows that followers of some members often do not follow much of the broader pack. Packs can therefore function as pathways to related accounts from the same niche that users may not encounter from their existing follow networks alone. Thus, pack fullness captures the potential of curated collections to support discovery by connecting users to socially and topically related accounts.

\section{Conclusion}
In this work, we studied semantic coherence, follow-network structure, and curator--member resemblance of Bluesky starter packs. We find that, within broad topic areas, starter packs capture semantically coherent subgroups of accounts. Pack members are more similar to their curators in posting content than comparable same-topic accounts, and demographic resemblance is positive for some attributes and topics. Shared followers and high reciprocity among pack members and between members and curators indicate that starter packs capture both social and topical structure.

However, the audience of an individual pack member typically follows only a small fraction of the remaining members. Therefore, even when starter pack members are topically and socially related, packs can expose users to accounts beyond their immediate follow networks. More broadly, these findings show that human-curated collections can represent meaningful community structure and can complement algorithmic approaches to discovery. Platforms could use such collections beyond initial onboarding and curator-shared links, for example by recommending relevant packs to users who already follow a part of a pack's members.

Future work can examine changes in follow relationships before and after pack inclusion, test whether recommending partially followed packs helps users discover relevant accounts, and assess whether these findings extend to less adopted starter packs. Comparisons with accounts outside starter packs could also help distinguishing pre-existing differences from changes associated with inclusion.

\bibliography{aaai2026}

\setcounter{figure}{0}
\setcounter{table}{0}

\renewcommand{\thefigure}{A.\arabic{figure}}
\renewcommand{\thetable}{A.\arabic{table}}

\begin{table*}[t]
\centering
\small
\caption{Topic composition, within-pack semantic similarity, and pack fullness. \emph{Packs} and \emph{Share} refer to counts in the analyzed corpus (\(N=2{,}718\)). The similarity measures are unweighted averages of the mean cosine similarity within each pack: across member pairs ($\text{cos}(u, v)$); between members and the pack curator ($\text{cos}(u, p_c)$); between members and the pack description ($\text{cos}(e_u, d_p)$); and between each member and the leave-one-out centroid of all other members ($\text{cos}(e_u, \mu_{p, -u})$). \emph{Fullness} is the share of packs within a topic with a ratio of \(0.5\) or less, calculated among followers $F_p$ that follow at least 5 members of the corresponding pack. Because similarity and fullness are calculated on filtered subsets, they do not necessarily cover the same packs counted in the \emph{Packs} column.}
\label{tab:topic_similarity_fullness}
\begin{tabular}{lrrccccc}
\toprule
Topic & Packs & Share & $\text{cos}(u, v)$ & $\text{cos}(u, c_p)$ & $\text{cos}(e_u, d_p)$ & $\text{cos}(e_u, \mu_{p, -u})$ & Fullness $\leq 0.5$ \\
\midrule
Academia \& Research          & 273 & 10.04\% & 0.8831 & 0.9097 & 0.7503 & 0.9377 & 0.8938 \\
Artists (Visual, SFW)         & 72  & 2.65\%  & 0.8972 & 0.9148 & 0.7497 & 0.9456 & 0.9306 \\
Authors \& Literary           & 92  & 3.38\%  & 0.9139 & 0.9270 & 0.7630 & 0.9547 & 0.9130 \\
Biology \& Life Sciences      & 95  & 3.50\%  & 0.8849 & 0.9162 & 0.7515 & 0.9391 & 0.9474 \\
Climate \& Environment        & 93  & 3.42\%  & 0.8954 & 0.9222 & 0.7586 & 0.9448 & 0.9462 \\
Educators (K--12)             & 82  & 3.02\%  & 0.8946 & 0.9201 & 0.7664 & 0.9433 & 0.7561 \\
Fan Communities               & 56  & 2.06\%  & 0.9119 & 0.9152 & 0.7478 & 0.9534 & 0.8571 \\
Foreign Language Communities  & 104 & 3.83\%  & 0.8935 & 0.9160 & 0.7448 & 0.9418 & 0.8039 \\
Gaming \& Game Development     & 81  & 2.98\%  & 0.9092 & 0.9241 & 0.7552 & 0.9521 & 0.9012 \\
LGBTQ+ Community              & 26  & 0.96\%  & 0.9213 & 0.9366 & 0.7688 & 0.9578 & 0.8462 \\
Medicine \& Healthcare        & 120 & 4.42\%  & 0.8744 & 0.9065 & 0.7504 & 0.9326 & 0.8667 \\
Music                         & 48  & 1.77\%  & 0.8852 & 0.9077 & 0.7224 & 0.9385 & 0.7917 \\
NSFW / Adult Content          & 213 & 7.84\%  & 0.9413 & 0.9492 & 0.8005 & 0.9687 & 0.8585 \\
Other / General               & 370 & 13.61\% & 0.9011 & 0.9202 & 0.7401 & 0.9468 & 0.8076 \\
Politics \& Journalism        & 482 & 17.73\% & 0.9224 & 0.9386 & 0.7492 & 0.9591 & 0.8838 \\
Sports                        & 186 & 6.84\%  & 0.9110 & 0.9118 & 0.7366 & 0.9527 & 0.8172 \\
Tech \& Software Development  & 325 & 11.96\% & 0.8905 & 0.9134 & 0.7693 & 0.9416 & 0.8462 \\
\bottomrule
\end{tabular}
\end{table*}

\begin{table*}[t]
\centering
\small
\setlength{\tabcolsep}{4pt}
\caption{Topic-level coverage of the demographic labels used in \textbf{RQ2}. Member and curator counts are unique within topic; an account associated with multiple topics is counted once in each of those topics. Consequently, the member and curator columns do not necessarily sum to the total row, which counts each account once across all topics. Median $|M_p|$ is the median number of members per pack. Median followers is calculated by first taking the median follower count among the members of each pack and then taking the median of these pack-level values within each topic. The \emph{skipped} columns report label coverage. For M3, skipped accounts are accounts that the model did not score. For ethnicolr, accounts are skipped if the model did not assign a probability of at least 0.5 to any label. Packs are skipped if either the curator or fewer than 10 non-curator members have a valid label.}
\label{tab:topic_coverage}

\begin{tabular}{lrrrrrrrrr}
\toprule
& & & & & & \multicolumn{2}{c}{M3 skipped}
& \multicolumn{2}{c}{ethnicolr skipped} \\
\cmidrule(lr){7-8}\cmidrule(lr){9-10}
Topic
& Packs
& Members
& \parbox{1cm}{\raggedleft Unique\\curators}
& Median $|M_p|$
& \parbox{1.5cm}{\raggedleft Median\\followers}
& Packs
& Members
& Packs
& Members \\
\midrule
Academia \& Research
& 273 & 18,335 & 262 & 65 & 677.00
& 129 & 7,922 & 161 & 10,192 \\

Artists (Visual, SFW)
& 72 & 5,393 & 71 & 70.5 & 628.75
& 36 & 2,696 & 46 & 3,713 \\

Authors \& Literary
& 92 & 6,815 & 88 & 74 & 990.00
& 51 & 3,069 & 59 & 4,073 \\

Biology \& Life Sciences
& 95 & 6,939 & 86 & 85 & 871.00
& 43 & 2,849 & 56 & 3,776 \\

Climate \& Environment
& 93 & 5,799 & 87 & 73 & 1,504.00
& 37 & 2,437 & 46 & 3,251 \\

Educators (K--12)
& 82 & 4,287 & 81 & 52.5 & 476.25
& 46 & 2,023 & 65 & 2,566 \\

Fan Communities
& 56 & 4,479 & 55 & 82 & 542.00
& 34 & 2,537 & 45 & 3,608 \\

Foreign Language Communities
& 104 & 5,190 & 101 & 41 & 1,317.00
& 66 & 2,944 & 88 & 3,973 \\

Gaming \& Game Development
& 81 & 5,875 & 75 & 72 & 1,097.00
& 50 & 2,903 & 54 & 4,127 \\

LGBTQ+ Community
& 26 & 1,309 & 26 & 40.5 & 2,692.00
& 15 & 734 & 22 & 1,028 \\

Medicine \& Healthcare
& 120 & 7,713 & 116 & 67.5 & 566.75
& 70 & 3,780 & 86 & 4,850 \\

Music
& 48 & 2,923 & 45 & 58 & 273.00
& 26 & 1,347 & 38 & 2,064 \\

NSFW / Adult Content
& 213 & 9,783 & 207 & 44 & 2,254.50
& 158 & 5,979 & 201 & 8,242 \\

Other / General
& 370 & 21,189 & 363 & 50 & 906.25
& 202 & 9,825 & 286 & 14,034 \\

Politics \& Journalism
& 482 & 25,658 & 452 & 63 & 2,805.50
& 227 & 11,200 & 338 & 15,441 \\

Sports
& 186 & 10,251 & 164 & 55.5 & 1,599.00
& 94 & 4,385 & 142 & 6,621 \\

Tech \& Software Development
& 325 & 16,279 & 309 & 55 & 788.00
& 152 & 6,842 & 211 & 9,545 \\

\midrule
\textbf{Total}
& \textbf{2,718}
& \textbf{144,987}
& \textbf{2,538}
& \textbf{59}
& \textbf{1,058.00}
& \textbf{1,436}
& \textbf{67,938}
& \textbf{1,944}
& \textbf{93,580} \\
\bottomrule
\end{tabular}
\end{table*}

\begin{table*}[t]
\centering
\caption{Demographic composition of curators and members by topic. Values are percentages. Race: W = White, B = Black,
A = Asian, H = Hispanic; gender: F = Female, M = Male. Member percentages use unique member accounts within each topic. \emph{Mean} averages the topic rows equally; \emph{Pooled} uses one denominator
over all packs or all unique members.}
\label{tab:demographic_composition}
\scriptsize
\setlength{\tabcolsep}{2.5pt}

\resizebox{\textwidth}{!}{%
\begin{tabular}{lrrrrrrrrrr@{\hspace{7pt}}rrrrrrrrrr}
\toprule
& \multicolumn{10}{c}{\textbf{Curator}}
& \multicolumn{10}{c}{\textbf{Member}} \\
\cmidrule(lr){2-11}
\cmidrule(lr){12-21}

& \multicolumn{4}{c}{Race}
& \multicolumn{4}{c}{Age}
& \multicolumn{2}{c}{Gender}
& \multicolumn{4}{c}{Race}
& \multicolumn{4}{c}{Age}
& \multicolumn{2}{c}{Gender} \\

\cmidrule(lr){2-5}
\cmidrule(lr){6-9}
\cmidrule(lr){10-11}
\cmidrule(lr){12-15}
\cmidrule(lr){16-19}
\cmidrule(lr){20-21}

\textbf{Topic}
& W & B & A & H
& $\leq$18 & 19--29 & 30--39 & $\geq$40
& F & M
& W & B & A & H
& $\leq$18 & 19--29 & 30--39 & $\geq$40
& F & M \\
\midrule

Academia \& Research
& 87.3 & 4.0 & 4.0 & 4.8
& 1.3 & 25.5 & 10.2 & 63.1
& 29.9 & 70.1
& 86.0 & 2.8 & 5.7 & 5.5
& 4.9 & 29.1 & 8.8 & 57.3
& 39.4 & 60.6 \\

Artists (Visual, SFW)
& 90.9 & 0.0 & 3.0 & 6.1
& 12.8 & 35.9 & 17.9 & 33.3
& 28.2 & 71.8
& 88.4 & 4.1 & 3.5 & 4.0
& 20.7 & 30.7 & 12.5 & 36.2
& 39.4 & 60.6 \\

Authors \& Literary
& 94.6 & 5.4 & 0.0 & 0.0
& 0.0 & 11.4 & 15.9 & 72.7
& 61.4 & 38.6
& 90.8 & 4.1 & 2.4 & 2.7
& 8.2 & 24.7 & 11.9 & 55.2
& 64.8 & 35.2 \\

Biology \& Life Sciences
& 76.2 & 2.4 & 4.8 & 16.7
& 1.8 & 21.4 & 12.5 & 64.3
& 19.6 & 80.4
& 83.7 & 1.7 & 5.8 & 8.7
& 5.8 & 35.7 & 9.8 & 48.7
& 33.7 & 66.3 \\

Climate \& Environment
& 82.7 & 7.7 & 5.8 & 3.8
& 8.2 & 14.8 & 9.8 & 67.2
& 34.4 & 65.6
& 86.7 & 5.1 & 2.5 & 5.7
& 4.8 & 22.5 & 7.8 & 65.0
& 33.4 & 66.6 \\

Educators (K--12)
& 95.8 & 4.2 & 0.0 & 0.0
& 6.5 & 17.4 & 15.2 & 60.9
& 45.7 & 54.3
& 92.4 & 2.6 & 1.7 & 3.3
& 6.5 & 24.8 & 13.6 & 55.0
& 56.2 & 43.8 \\

Fan Communities
& 69.2 & 0.0 & 15.4 & 15.4
& 33.3 & 25.0 & 12.5 & 29.2
& 45.8 & 54.2
& 85.3 & 5.5 & 4.9 & 4.2
& 32.8 & 30.5 & 14.7 & 21.9
& 48.9 & 51.1 \\

Foreign Language Communities
& 66.7 & 8.3 & 12.5 & 12.5
& 22.9 & 25.0 & 14.6 & 37.5
& 18.8 & 81.2
& 83.1 & 4.7 & 3.1 & 9.1
& 21.9 & 26.8 & 8.3 & 43.0
& 29.4 & 70.6 \\

Gaming \& Game Development
& 87.1 & 6.5 & 0.0 & 6.5
& 24.2 & 27.3 & 15.2 & 33.3
& 12.1 & 87.9
& 86.7 & 6.4 & 2.8 & 4.2
& 30.2 & 29.4 & 19.3 & 21.1
& 16.4 & 83.6 \\

LGBTQ+ Community
& 85.7 & 0.0 & 0.0 & 14.3
& 14.3 & 35.7 & 7.1 & 42.9
& 42.9 & 57.1
& 85.1 & 7.8 & 3.9 & 3.2
& 21.9 & 42.4 & 13.6 & 22.1
& 40.7 & 59.3 \\

Medicine \& Healthcare
& 89.4 & 2.1 & 4.3 & 4.3
& 5.0 & 18.3 & 11.7 & 65.0
& 28.3 & 71.7
& 85.5 & 3.6 & 6.2 & 4.6
& 5.6 & 26.6 & 10.5 & 57.3
& 46.9 & 53.1 \\

Music
& 82.4 & 17.6 & 0.0 & 0.0
& 23.1 & 11.5 & 23.1 & 42.3
& 11.5 & 88.5
& 89.2 & 6.1 & 2.6 & 2.2
& 22.4 & 25.4 & 12.2 & 40.0
& 25.4 & 74.6 \\

NSFW / Adult Content
& 76.2 & 0.0 & 19.0 & 4.8
& 26.8 & 32.4 & 26.8 & 14.1
& 39.4 & 60.6
& 79.9 & 9.0 & 6.3 & 4.8
& 29.4 & 39.2 & 17.0 & 14.4
& 38.5 & 61.5 \\

Other / General
& 93.0 & 4.4 & 0.9 & 1.8
& 13.2 & 28.4 & 8.8 & 49.5
& 31.4 & 68.6
& 88.6 & 5.5 & 2.4 & 3.5
& 14.6 & 25.4 & 11.9 & 48.1
& 40.8 & 59.2 \\

Politics \& Journalism
& 90.4 & 5.6 & 1.7 & 2.3
& 9.3 & 20.4 & 11.1 & 59.3
& 26.1 & 73.9
& 88.7 & 4.6 & 2.4 & 4.3
& 9.5 & 24.6 & 9.7 & 56.2
& 38.2 & 61.8 \\

Sports
& 89.3 & 5.4 & 3.6 & 1.8
& 27.5 & 26.7 & 10.8 & 35.0
& 26.7 & 73.3
& 92.2 & 3.3 & 2.3 & 2.3
& 19.6 & 30.4 & 14.6 & 35.4
& 13.6 & 86.4 \\

Tech \& Software Dev
& 92.6 & 0.7 & 3.7 & 2.9
& 8.9 & 23.2 & 18.4 & 49.5
& 10.5 & 89.5
& 87.7 & 2.1 & 5.2 & 5.0
& 12.0 & 33.1 & 16.6 & 38.3
& 19.0 & 81.0 \\

\midrule
\textbf{Mean of topics}
& \textbf{85.3} & \textbf{4.4} & \textbf{4.6} & \textbf{5.7}
& \textbf{14.1} & \textbf{23.5} & \textbf{14.2} & \textbf{48.2}
& \textbf{30.2} & \textbf{69.8}
& \textbf{87.1} & \textbf{4.6} & \textbf{3.7} & \textbf{4.6}
& \textbf{15.9} & \textbf{29.5} & \textbf{12.5} & \textbf{42.1}
& \textbf{36.7} & \textbf{63.3} \\

\textbf{Pooled}
& \textbf{88.3} & \textbf{4.2} & \textbf{3.4} & \textbf{4.1}
& \textbf{11.9} & \textbf{23.6} & \textbf{13.2} & \textbf{51.2}
& \textbf{27.5} & \textbf{72.5}
& \textbf{87.5} & \textbf{4.0} & \textbf{3.8} & \textbf{4.7}
& \textbf{13.5} & \textbf{29.0} & \textbf{12.2} & \textbf{45.4}
& \textbf{35.7} & \textbf{64.3} \\
\bottomrule

\end{tabular}%
}

\vspace{2pt}

\end{table*}

\begin{table*}[t]
\centering
\caption{Cosine similarity by topic for the same sampled account pairs. Full-history embeddings use all available posts, whereas originals-only embeddings use standalone posts authored by each account, excluding reposts, replies, and quote posts.}
\small
\begin{tabular}{lrrrrrrrr}
\toprule
& \multicolumn{4}{c}{Same-topic pairs} & \multicolumn{4}{c}{Same-pack pairs} \\
\cmidrule(lr){2-5}\cmidrule(lr){6-9}
Topic & Full mean & Orig. mean & Full med. & Orig. med. & Full mean & Orig. mean & Full med. & Orig. med. \\
\midrule
Academia \& Research & 0.8254 & 0.7817 & 0.8322 & 0.7881 & 0.8973 & 0.8535 & 0.9116 & 0.8700 \\
Artists (Visual, SFW) & 0.8388 & 0.8177 & 0.8535 & 0.8313 & 0.9096 & 0.8890 & 0.9276 & 0.9077 \\
Authors \& Literary & 0.8746 & 0.8529 & 0.8912 & 0.8704 & 0.9227 & 0.9012 & 0.9371 & 0.9166 \\
Biology \& Life Sciences & 0.8266 & 0.7590 & 0.8373 & 0.7685 & 0.8998 & 0.8318 & 0.9143 & 0.8485 \\
Climate \& Environment & 0.8306 & 0.7943 & 0.8423 & 0.8017 & 0.9002 & 0.8661 & 0.9153 & 0.8820 \\
Educators (K-12) & 0.8277 & 0.7981 & 0.8407 & 0.8086 & 0.9043 & 0.8717 & 0.9193 & 0.8881 \\
Fan Communities & 0.8556 & 0.8370 & 0.8767 & 0.8576 & 0.9243 & 0.9010 & 0.9457 & 0.9229 \\
Foreign Language Comm. & 0.7338 & 0.7104 & 0.7294 & 0.7049 & 0.9094 & 0.8937 & 0.9327 & 0.9198 \\
Gaming and Game Dev. & 0.8723 & 0.8489 & 0.8864 & 0.8630 & 0.9182 & 0.8948 & 0.9345 & 0.9119 \\
LGBTQ+ Community & 0.8658 & 0.8489 & 0.8803 & 0.8632 & 0.9417 & 0.9268 & 0.9592 & 0.9486 \\
Medicine \& Healthcare & 0.8237 & 0.7767 & 0.8301 & 0.7840 & 0.8940 & 0.8458 & 0.9086 & 0.8615 \\
Music & 0.8374 & 0.8208 & 0.8485 & 0.8308 & 0.8945 & 0.8788 & 0.9131 & 0.8973 \\
NSFW / Adult Content & 0.8961 & 0.8788 & 0.9096 & 0.8936 & 0.9465 & 0.9282 & 0.9600 & 0.9430 \\
Other / General & 0.8452 & 0.8254 & 0.8629 & 0.8403 & 0.9163 & 0.8972 & 0.9347 & 0.9164 \\
Politics \& Journalism & 0.8203 & 0.7885 & 0.8266 & 0.7898 & 0.9265 & 0.9022 & 0.9426 & 0.9194 \\
Sports & 0.8117 & 0.7988 & 0.8186 & 0.8029 & 0.9150 & 0.9025 & 0.9329 & 0.9229 \\
Tech \& Software Dev & 0.8387 & 0.8119 & 0.8477 & 0.8212 & 0.9046 & 0.8750 & 0.9199 & 0.8942 \\
\bottomrule
\end{tabular}
\label{tab:cosine-full-vs-originals-by-topic}
\end{table*}

\begin{table*}[t]
\centering
\caption{
Mean and median cosine similarity between member pairs using full account embeddings and originals-only account embeddings. Total embeddings incorporate original posts, reposts, and quote posts, whereas originals-only embeddings use only standalone posts authored by each account, excluding reposts, replies, and quote posts.
}
\label{tab:cosine-total-originals}
\begin{tabular}{lrrrr}
\toprule
Distribution
& \multicolumn{2}{c}{Total}
& \multicolumn{2}{c}{Originals only} \\
\cmidrule(lr){2-3}
\cmidrule(lr){4-5}
& Mean & Median & Mean & Median \\
\midrule
Random pairs     & 0.7885 & 0.7944 & 0.7674 & 0.7721 \\
Same-topic pairs & 0.8320 & 0.8462 & 0.8088 & 0.8199 \\
Same-pack pairs  & 0.9091 & 0.9277 & 0.8858 & 0.9051 \\
\bottomrule
\end{tabular}
\end{table*}

\begin{figure*}[!t]
\begin{promptbox}[label={prompt:topics}]{Starter pack topic detection template}
You are a classifier for Bluesky starter packs — curated lists of accounts people recommend to follow.

Given a starter pack name and description, assign it to the most relevant topics from the list below.

\vspace{0.5em}
\#\# Topics

- Politics \& Journalism: Political figures, journalists, reporters, media organizations, news commentary

- Artists (Visual, SFW): Visual artists, illustrators, painters, animators, comic artists (non-adult)

- NSFW / Adult Content: Adult content creators, explicit material, OnlyFans, sex workers

- LGBTQ+ Community: Queer, gay, lesbian, trans, nonbinary, pride communities

- Sports: Sports teams, athletes, fans, fantasy sports, specific leagues

- Authors \& Literary: Writers, novelists, poets, publishers, literary agents, book communities

- Game Development: Game developers, designers, studios, indie games, tabletop RPGs, board games

- Climate \& Environment: Climate scientists, environmentalists, conservation, sustainability, energy

- Music: Musicians, bands, singers, composers, DJs, music producers

- Educators (K-12): Teachers, librarians, classroom educators, school communities

- Academia \& Research: University faculty, researchers, PhD students, academic institutions

- Biology \& Life Sciences: Biologists, genomics, neuroscience, genetics, bioinformatics

- Medicine \& Healthcare: Doctors, nurses, clinicians, medical specialties, healthcare

- Tech \& Software Dev: Software developers, engineers, cybersecurity, AI/ML, open source

- Foreign Language Communities: Non-English communities (French, Portuguese, German, Spanish, Korean, etc.)

- Fan Communities: K-pop fans, anime, manga, fandom, stan communities

- Other / General: Does not fit any category above

\vspace{0.5em}
\#\# Instructions

- Return ONLY a valid JSON object, no explanation outside the JSON.

- Assign up to 3 topics in order of relevance.

- If only 1 or 2 topics clearly apply, set lower-ranked confidences to 0.0 and use "Other / General".

- Confidence is a float between 0.0 and 1.0.

- Reasoning is one sentence explaining the primary assignment only.

- Description may be empty — classify on name alone in that case.

\vspace{0.5em}
Name: \pvar{PACK\_TITLE}

Description: \pvar{PACK\_DESCRIPTION}

\vspace{0.5em}
\#\# Output

\begin{lstlisting}[
  basicstyle=\footnotesize\ttfamily,
  frame=none,
  numbers=none,
  xleftmargin=1.5em
]
{
  "primary": "<label>",
  "primary_confidence": <float>,
  "second": "<label>",
  "second_confidence": <float>,
  "third": "<label>",
  "third_confidence": <float>,
  "reasoning": "<one sentence>"
}
\end{lstlisting}
\end{promptbox}
\end{figure*}

\end{document}